\documentclass[12pt]{iopart}

\usepackage{iopams}  
\begin{document}

\title{Renormalization group in difference systems}

\author{M Iwasa and K Nozaki}

\address{Department of Physics, Nagoya university,
Nagoya 464-8602, Japan}
\ead{miwasa@r.phys.nagoya-u.ac.jp}
\begin{abstract}\\
 A new singular perturbation method based on the Lie symmetry group is
 presented to a system of difference equations.
 This method yields consistent derivation of a renormalization group 
 equation which gives an asymptotic solution of the difference equation. 
 The renormalization group equation is a Lie differential equation of 
 a Lie group which leaves the system approximately invariant.
 For a 2-D symplectic map,  the renormalization group equation becomes a
 Hamiltonian system and a long-time behaviour of the symplectic map is
 described by the Hamiltonian.  We study the Poincar\'e-Birkoff 
 bifurcation in the 2-D symplectic map by means of the Hamiltonian and
 give a condition for the  bifurcation.  
\end{abstract}
\section{\label{sec:introduction}Introduction} 
  There have been many studies concerning application of renormalization
  group method of quantum field theory as a singular perturbation method
  to treat differential equations since the work of Illinois groups
  \cite{ChenGoldenfeldOono}\cite{GotoMasutomiNozaki}.     
  Although this method enables us to remove secular or divergent terms
  appearing in a naive perturbation solution by renormalizing integral
  constants appearing in the lowest-order of perturbation solution and
  to give a well-behaved asymptotic solution, it is necessary to
  calculate a naive perturbation solution to derive the renormalization
  group equation, and whether the system could be renormalized or not
  depends on the functional form of a naive perturbation solution.  
  The present authors presented another renormalization group method in
  terms of the Lie symmetry group \cite{IwasaNozaki}. 
  Without calculating a naive perturbation solution, the renormalized
  solution is constructed as an invariant solution under transformation
  which leaves the system approximately invariant. 
  The approximate Lie group transforms an unperturbed  solution into a
  renormalized perturbed solution.  
  The both renormalization group methods were based on the theory of
  continuous group and so they were developed to apply a system of
  differential equations. 
  For some discrete dynamical systems such as a system of difference
  equations, there has been found a working procedure to derive a
  renormalized difference equation by means of a naive perturbation
  solution \cite{MaruoGotoNozaki}\cite{KanoNozaki}.  
  However, the concept of group was not used in the procedure  to derive a
  renormalized discrete system and so legitimacy of the result is not
  clear. 
  The purpose of this paper is to develop a renormalization group method
  in terms of the Lie symmetry group for  a system of difference
  equations. Apparently, it seems to be difficult to use the same
  procedure as in the case of a differential equation when we treat a
  discrete system because Lie symmetry groups are one-parameter
  continuous groups. 
  However, we regard a discrete system as an
  algebraic system and consider a particular class of approximate Lie
  symmetry groups which transforms a perturbed system to an unperturbed
  system approximately. Then we succeed in obtaining a renormalized
  perturbed solution by means of the so-called Lie differential equation
  corresponding to the approximate Lie symmetry. 
  Thus, we present a renormalization group method for a system of
  difference equations, where the continuous group theory is fully
  employed, and the Lie equation becomes the renormalization group
  equation for a difference system.   
  As an application of the new method, we study the Poincar\'e-Birkoff
  bifurcation in a 2-D symplectic map and give a condition for the
  bifurcation explicitly. 
 
\section{\label{sec:RGmethod}Renormalization group method with Lie
symmetry}
Let us consider the following 2-D symplectic map of action-angle type, $(u_n,v_n)\mapsto(u_{n+1},v_{n+1})$:
\begin{eqnarray}
	 u_{n+1}=u_n+v_{n+1}\nonumber,\\
         v_{n+1}=v_n+au_n+\varepsilon g(u_n),
 \label{eq:2-1}
\end{eqnarray}
where the coefficient $a \in \mathbb{R}$ is constant, $g$ is a function
of $u_n$, and $\varepsilon$ is a perturbation parameter which is
small.
The system (\ref{eq:2-1}) reads 
\begin{eqnarray}
 \left(
   \begin{array}{c}
    u_{n+1}\\
    v_{n+1}
   \end{array}
 \right)  
  =
 \left(
   \begin{array}{cc}
    a+1&1 \\
    a  &1 
   \end{array}
 \right) 
 \left(
  \begin{array}{c}
  u_n\\
  v_n
  \end{array}
 \right)
  +
 \left(
  \begin{array}{c}
   \varepsilon g(u_n)\\
   \varepsilon g(u_n)
  \end{array}
 \right).
\label{eq:2-2}
\end{eqnarray}
Under a transformation of coordinates $(u_n,v_n)\mapsto
(z_n,\overline{z_n})$ such as
   \begin{eqnarray}
    \left(
      \begin{array}{c}
        z_n\\
        \overline{z_n}
      \end{array} 
    \right) 
     =
    \left(
      \begin{array}{cc}
        \frac{1}{2\cos (\omega/2)}&
	\rmi\frac{\exp{(\rmi\omega/2)}}{2\sin\omega\cos\omega}\\
        \frac{1}{2\cos (\omega/2)}&
        -\rmi\frac{\exp{(-\rmi\omega/2)}}{2\sin\omega \cos \omega}
      \end{array}
    \right)
    \left(
      \begin{array}{c}
        u_n\\
        v_n
      \end{array}
    \right),
\label{eq:2-3}\\
    \Longleftrightarrow 
    \left(
      \begin{array}{c}
        u_n\\
        v_n  
      \end{array}
    \right)
    =
    \left(
      \begin{array}{cc}
        \exp{(-\rmi\omega/2)}  & \exp{(\rmi\omega/2)} \\
        -2\rmi\sin(\omega/2)   & 2\rmi\sin(\omega/2)
      \end{array}
    \right)
    \left(
      \begin{array}{c}
        z_n\\
        \overline{z_n}
      \end{array} 
    \right),
    \label{eq:2-4}
\end{eqnarray}
the linear part of (\ref{eq:2-2}) is diagonalized as follows: 
\begin{eqnarray}
    \left(
      \begin{array}{c}
       z_{n+1}\\
       \overline{z_{n+1}}
      \end{array}
    \right)
     =
    \left(
      \begin{array}{cc}
       \rme^{-\rmi\omega}&0     \\
       0                 &\rme^{\rmi\omega}
      \end{array}
    \right)
    \left(
      \begin{array}{c}
       z_n\\
       \overline{z_n}
      \end{array}
    \right)
    +\varepsilon 
     \frac{g(u_n)}{2\sin\omega}
     \left(
      \begin{array}{c}
       \rmi\exp{(-i\omega/2)}\\
       -\rmi\exp{(i\omega/2)}
      \end{array}
     \right),
    \label{eq:2-5}
\end{eqnarray}
where $a+2=:2\cos \omega$, and $\overline{z_n}$ denotes
the complex conjugate of $z_n$. 
Here $0 < a < -4$ because we are interested in a case the origin
is assumed to be elliptic.

Since the second component of (\ref{eq:2-5}) is the complex
conjugate of the first, we have only to consider the first component of
the equation, which is 
\begin{eqnarray}
 z_{n+1}=\rme^{-\rmi\omega}z_n+\varepsilon
  \frac{\rmi\exp{(-\rmi\omega/2})}{2\sin \omega}g(u_n).
\label{eq:2-6}
\end{eqnarray}    

Let (\ref{eq:2-6}) admit a Lie group transformation whose
infinitesimal generator takes the form
\begin{eqnarray}
 \fl X(n,z_n,\overline{z_n},z_{n+1},\overline{z_{n+1}})
     = \partial_\varepsilon+\eta^z(n,z_n,\overline{z_n})\partial_{z_n}
     +\eta^{\overline z}(n,z_n,\overline{z_n})\partial_{\overline{z_n}} 
     \nonumber\\
     +\eta^z(n+1,z_{n+1},\overline{z_{n+1}})\partial_{z_{n+1}}
     +\eta^{\overline
     z}(n+1,z_{n+1},\overline{z_{n+1}})\partial_{\overline{z_{n+1}}}.
\label{eq:2-7}
\end{eqnarray}
We can rewrite (\ref{eq:2-7}) to
\begin{eqnarray}
 \fl X(n,z_n,\overline{z_n},z_{n+1},\overline{z_{n+1}})
     = \partial_\varepsilon+\eta(n,z_n,\overline{z_n})\partial_{z_n}
     +\overline{\eta(n,z_n,\overline{z_n})}\partial_{\overline{z_n}} 
     \nonumber\\
     +\eta(n+1,z_{n+1},\overline{z_{n+1}})\partial_{z_{n+1}}
     +\overline{\eta(n+1,z_{n+1},\overline{z_{n+1}})}\partial_{\overline{z_{n+1}}},
\label{eq:2-8}
\end{eqnarray}
because it can be shown that $\eta^{\overline z}=\overline{\eta^z}$.
Then the determining equation for (\ref{eq:2-6}), which determines
$\eta(n,z_n,\overline{z_n})$ of the vector field (\ref{eq:2-8}) \cite{Olver}, is
given by 
\begin{eqnarray}
\fl
     X(n,z_n,\overline{z_n},z_{n+1},\overline{z_{n+1}})
     \left\{
       z_{n+1}-\rme^{-\rmi\omega}z_n
       -\varepsilon \frac{\rmi\exp{(-\rmi\omega/2)}}
                         {2\sin \omega}
                    g(u_n)
     \right\}\Biggm|_{
         {\rm \Eref{eq:2-6}}}\hspace{-0.6cm}=0.
\label{eq:2-9}
\end{eqnarray}  
Because we wish to find such a symmetry that leaves the system
approximately invariant to leading order, we
need only to solve the following leading-order determining equation:
\begin{eqnarray}
\fl      
    X(n,z_n,\overline{z_n},z_{n+1},\overline{z_{n+1}})
        \left\{
          z_{n+1}-\rme^{-\rmi\omega}z_n
          -\varepsilon 
            \frac{\rmi\exp{(-\rmi\omega/2)}}
                 {2\sin\omega}
           g(u_n)
        \right\}\Biggm|_{
         {\scriptsize z_{n+1}=\rme^{-\rmi\omega}z_n }}
        \hspace{-1.5cm}=\Or(\varepsilon),\\
    \Longleftrightarrow 
         \eta(n+1,\rme^{-\rmi\omega}z_n,\rme^{\rmi\omega}\overline{z_n})
         -\rme^{-\rmi\omega}\eta(n,z_n,\overline{z_n})
         =\frac{\rmi\exp{(-\rmi\omega/2)}}
               {2\sin \omega}
          g(u_n).
\label{eq:2-10}
\end{eqnarray}
By solving (\ref{eq:2-10}), we obtain the infinitesimal generator $X$
admitted by the system (\ref{eq:2-6}) in the leading-order approximation. 

Using the obtained infinitesimal generator $X$, we construct a group-invariant
solution of the system (\ref{eq:2-6}).
The group-invariant solution, $z_n = z_n(\varepsilon)$,
satisfies the following Lie equation: 
\begin{eqnarray}
 X\left\{z_n-z_n(\varepsilon)\right\}\bigl|_{z_n=z_n(\varepsilon)}=0,
\label{eq:2-11}
\end{eqnarray} 
which reads
\begin{eqnarray}
 \frac{\rmd z_n(\varepsilon)}{\rmd\varepsilon} 
  = \eta(n,z_n(\varepsilon),\overline{z_n(\varepsilon)}).
\label{eq:2-12}
\end{eqnarray}
We refer to the Lie equation (\ref{eq:2-12}) as the renormalization group
equation.
Solving the renormalization group equation by adopting solutions of
unperturbed system as a boundary condition, i.e.
\begin{eqnarray}
 z_n(\varepsilon=0)=z_n^{(0)},
\label{eq:2-13}
\end{eqnarray}
where $z_n=z_n^{(0)}$ denotes the solution of the unperturbed system,
we obtain an asymptotic solution of the system.

\section{\label{power series}Power series nonlinear terms}
Let us consider a case $g(u_n)$ is a power series
with respect to $u_n$, i.e. we set 
\begin{eqnarray}
 g(u_n):= \sum_{j=0}^{\infty}A_{j}u_n^j.     
\label{eq:3-1}
\end{eqnarray}
where $A_{j}\in\mathbb{R}$. 
Under the linear transformation (\ref{eq:2-4}), 
(\ref{eq:3-1}) can be written as
\begin{eqnarray}
 g(u_n(z_n,\overline{z_n}))=\sum_{l,m}B_{lm}(\omega)z_n^l\overline{z_n}^m,
\label{eq:3-2}
\end{eqnarray}
where
\begin{eqnarray}
 B_{lm}(\omega)=A_{l+m}\ _{l+m}C_l\ \exp{[-\rmi(l-m)\omega/2]}.
 \label{eq:3-2.5}
\end{eqnarray}
Here $_kC_r:=k!/[r!(k-r)!]$. 
Then the determining equation (\ref{eq:2-10}) becomes
\begin{eqnarray}
\fl
  \eta(n+1,e^{-\rmi\omega}z_n,e^{\rmi\omega}\overline{z_n})
    -e^{-\rmi\omega}\eta(n,z_n,\overline{z_n})
   =\frac{\rmi\exp{(-\rmi\omega/2)}}
         {2\sin \omega}
    \sum_{l,m}B_{lm}(\omega)z_n^l\overline{z_n}^m.
\label{eq:3-3}
\end{eqnarray}
The nature of the solution of (\ref{eq:3-3}) depends on whether
$\omega/2\pi$ is rational or irrational. 

\subsection{$\omega/2\pi$ is irrational}
Let $\omega/2\pi$ be irrational,
then only terms proportional to $z_n^{l+1}\overline{z_n}^l$ cause
resonance, and the solution of the determining equation
(\ref{eq:3-3}) becomes  
\begin{eqnarray}
\fl
 \eta(n,z_n,{\overline z_n})
  =&\sum_{l,m\  {\rm s.t.}\ l-m=1}
    \frac{\rmi\exp{(\rmi\omega/2)}B_{lm}(\omega)}{2\sin\omega}
    nz_n^l\overline{z_n}^m
     \nonumber \\
\fl
   &+\sum_{l,m\  {\rm s.t.}\ l-m\neq1}
    \frac{\rmi\exp{(-\rmi\omega/2)}B_{lm}(\omega)}
         {2\sin\omega
           \left\{
             \exp{[-\rmi(l-m)\omega]}-\exp{[-\rmi\omega]}
           \right\}
         }
    z_n^l\overline{z_n}^m.
\label{eq:3-7}
\end{eqnarray}
Here $\exp{[-\rmi(l-m)\omega]}-\exp{[-\rmi\omega]}\neq 0$ for $l-m\neq 1$.
The terms proportional to $n$ are called secular terms.
The renormalization group equation i.e. (\ref{eq:2-12}) becomes 
\begin{eqnarray}
 \frac{\rmd z_n}{\rmd \varepsilon} 
  =&\sum_{l,m\  {\rm s.t.}\ l-m=1}
       \frac{\rmi\exp{(\rmi\omega/2)}B_{lm}(\omega)}
            {2\sin\omega}
        nz_n^l\overline{z_n}^m
     \nonumber \\
   &+\sum_{l,m\  {\rm s.t.}\ l-m\neq1}
       \frac{\rmi\exp{(\rmi\omega/2)}B_{lm}(\omega)}
            {2\sin\omega
               \left\{
                 \exp{[-\rmi(l-m)\omega]}-\exp{[-\rmi\omega}]
               \right\}
            }
        z_n^l\overline{z_n}^m.
\label{eq:3-8}
\end{eqnarray}
Because we are interested in a long-time behaviour of the system,
we consider the case of $n\gg1$.
Then we neglect non-secular terms and the renormalization group equation
becomes 
\begin{eqnarray}
  \frac{\rmd z_n}{\rmd\varepsilon} 
  &=\frac{\rmi\exp{(\rmi\omega/2)}}{2\sin\omega}
     \sum_{l=0}^{\infty}
     B_{l+1\ l}(\omega)
     n|z_n|^{2l}z_n,
  \nonumber \\
  &=\frac{\rmi}{2\sin\omega}
    \sum_{l=0}^{\infty}
    A_{2l+1}\ _{2l+1}C_ln|z_n|^{2l}z_n,
\label{eq:3-9}
\end{eqnarray}
where we use (\ref{eq:3-2.5}). 
Because $\rmd |z_n|/\rmd\varepsilon=0$, introducing $R:=|z_n|$, 
the solution of (\ref{eq:3-9}) becomes
\begin{eqnarray}
  z_n(\varepsilon)=z_n(0)
                   \exp{\left(
                          \frac{\rmi}{2\sin\omega}
                          \sum_{l=0}^{\infty}
                          A_{2l+1}\ _{2l+1}C_lR^{2l}n\varepsilon
                        \right)}.
\end{eqnarray} 
Thus, we obtain an asymptotic behaviour of a difference equation by
solving a differential equation.

\subsection{$\omega/2\pi$ is rational}
Let $\omega/2\pi$ be rational i.e. $\omega=2\pi q/p$ where $q,p \in
\mathbb{N}$,     
then denominators of the second terms in the right hand side of 
(\ref{eq:3-8}), $\exp{[-\rmi(l-m)\omega]}-\exp{[-\rmi\omega]}$, 
become zero when $l-m=1+kp$ for $k\in\mathbb{Z}$.
Therefore terms $z_n^l{\overline z_n}^m$ where $l-m=1+kp$ give secular
contribution to the determining equation (\ref{eq:3-3}). 
The solution of the determining equation becomes
\begin{eqnarray}
\fl
 \eta(n,z_n,{\overline z_n})
  =&\sum_{l,m\  {\rm s.t.}\ l-m=1+kp}
      \frac{\rmi\exp{(\rmi\omega/2)}B_{lm}(\omega)}{2\sin\omega}
     nz_n^l\overline{z_n}^m
  \nonumber\\
   &+\sum_{l,m\  {\rm s.t.}\l-m\neq1+kp}
     \frac{\rmi\exp{(-\rmi\omega/2)}B_{lm}(\omega)}
          {2\sin\omega
            \left\{
              \exp{[-\rmi(l-m)\omega]}-\exp{[-\rmi\omega]}
            \right\}
          }
     z_n^l\overline{z_n}^m.
\label{eq:3-10}
\end{eqnarray} 
In the case of $n\gg1$, the renormalization group equation reads
\begin{eqnarray}
 \frac{\rmd z_n}{\rmd \varepsilon}
   =\sum_{l,m\ {\rm s.t.}\ l-m=1+kp}
    \frac{\rmi\exp{(\rmi\omega/2)}B_{lm}(\omega)}{2\sin\omega}
    nz_n^l\overline{z_n}^m,
\label{eq:3-11}
\end{eqnarray}
which is rewritten as 
\begin{eqnarray}
  \frac{\rmd z_n}{\rmd \varepsilon}
    =\frac{\rmi\exp{(\rmi\omega/2)}n}{2\sin\omega}
      \Biggl[
             &\sum_{l\geq1,m\geq0}
            \hspace{-3mm} 
             B_{l+mp\ l-1}(\omega)
              |z_n|^{2(l-1)}z_n^{mp+1}
 \nonumber \\
             &+\sum_{l\geq0,m\geq1}
            \hspace{-3mm} 
             B_{l\ l+mp-1}(\omega)
              |z_n|^{2l}\overline{z_n}^{mp-1}
      \Biggr],
\label{eq:3-12}
\end{eqnarray}
or 
\begin{eqnarray}
\fl 
   \frac{\rmd z_n}{\rmd \tau}
    &=&\frac{\rmi\exp{(\rmi\omega/2)}}{2\sin\omega}
      \left[
            \sum_{l\geq1,m\geq0}
            \hspace{-3mm} 
             B_{l+mp\ l-1}(\omega)
              |z_n|^{2(l-1)}z_n^{mp+1}
             +\sum_{l\geq0,m\geq1}
            \hspace{-3mm} 
             B_{l\ l+mp-1}(\omega)
              |z_n|^{2l}\overline{z_n}^{mp-1}
      \right],
    \nonumber \\
\fl   &=&\frac{\rmi}{2\sin\omega}
        \Biggl[
          \sum_{l\geq1,m\geq0}
            A_{2l+mp-1}\ _{2l+mp-1}C_{l+mp}
            |z_n|^{2(l-1)}z_n^{mp+1}
        \nonumber \\
\fl   && 
\hspace{2cm} 
       +\sum_{l\geq0,m\geq1}
           A_{2l+mp-1}\ _{2l+mp-1}C_l
              |z_n|^{2l}\overline{z_n}^{mp-1}
        \Biggr],
\label{eq:3-12.5}
\end{eqnarray}
where $\tau:=\varepsilon n$.
This is a Hamiltonian system whose Hamiltonian is
\begin{eqnarray}
\fl H(z_n,\overline{z_n})
    =\sum_{\scriptsize
            {
              \begin{array}{c}
	      l\geq0,m\geq0\\
              (l,m)\neq(0,0)
	      \end{array}
            }
          }
            \left[
            \rmi A_{2l+mp-1} 
            \frac{(2l+mp-1)!}{(l+mp)!\ l!}
            |z_n|^{2l}\left(z_n^{mp}+\overline{z_n}^{mp}\right)
            \right].
\label{eq:3-12.7}
\end{eqnarray}
Because (\ref{eq:3-12.5}) describes also $n$ dependence of
$z_n(\varepsilon)$, the phase space structure of this Hamiltonian system
provides a long-time behaviour of the difference equation.  

\subsection{$\omega/2\pi$ is close to a rational number}
In order to study the Poincar\'e-Birkoff bifurcation in the symplectic
map, let us consider the case $\omega/2\pi$ is close to a rational number, 
\begin{eqnarray}
 \omega = \omega_0 + \varepsilon\delta,
\label{eq:3-13}
\end{eqnarray}
where $\omega_0/2\pi$ is a rational number, $\omega_0=2\pi q/p$.
The difference equation (\ref{eq:2-6}) becomes
\begin{eqnarray}
 z_{n+1}=e^{-\rmi\omega_0}z_n+\varepsilon
  \left[
    \frac{\rmi\exp{(-\rmi\omega_0/2)}}{2\sin \omega_0}g(u_n)
    -\rmi\delta z_n
  \right]+\Or(\varepsilon^2) .
\label{eq:3-14}
\end{eqnarray}  

With the same procedure as we have followed in the previous subsections,  
we can derive the renormalization group equation in the case of $n\gg1$
as follows:
\begin{eqnarray}
 \frac{\rmd z_n}{\rmd \tau}
      &=&\rmi\frac{1}{2\sin\omega_0}
        \Biggl[
          \sum_{l\geq1,m\geq0}
            A_{2l+mp-1}\ _{2l+mp-1}C_{l+mp}
            |z_n|^{2(l-1)}z_n^{mp+1}
        \nonumber \\
   && 
\hspace{2cm} 
       +\sum_{l\geq0,m\geq1}
           A_{2l+mp-1}\ _{2l+mp-1}C_l
              |z_n|^{2l}\overline{z_n}^{mp-1}
        \Biggr] 
        -\rmi\delta z_n.
\label{eq:3-15}
\end{eqnarray}

For simplicity, let us consider such a case as follows:
\begin{eqnarray}
 g(u_n)&:=&bu_n^{p-1},
             \nonumber\\
        &=&b\sum_{l=0}^{p-1}
               \ _{p-1}C_l
                 \exp{[-\rmi(2l-p+1)\omega/2]}
                 z_n^l \overline{z_n}^{p-l-1},
\label{eq:3-16}
\end{eqnarray}
where $b$ is a parameter.
Then the renormalization group equation (\ref{eq:3-15}), which
describes a long-time behaviour, becomes  
\begin{eqnarray}
  \frac{\rmd z_n(\tau)}{\rmd \tau}
   = \alpha\frac{\rmi b}{2\sin\omega_0}
     \ _{p-1}C_\frac{p}{2}
    z_n^{\frac{p}{2}}\overline{z_n}^{\frac{p}{2}-1}
  +\frac{\rmi b}{2\sin\omega_0}\overline{z_n}^{p-1}
  -\rmi\delta z_n,
   \label{eq:3-18}
\end{eqnarray} 
where $\alpha=1$ when $p$ is an even and $\alpha=0$ when $p$ is an odd.
This is a Hamiltonian system whose Hamiltonian is 
\begin{eqnarray}
  H= \alpha\frac{\rmi b}{p\sin\omega_0}
     \ _{p-1}C_\frac{p}{2}|z_n|^p
    +\frac{\rmi b}{2p\sin\omega_0}\left(z_n^p+\overline{z_n}^p\right)
    -\rmi\delta |z_n|^2.
   \label{eq:3-19}
\end{eqnarray}
At fixed points $z_n$ satisfies
\begin{eqnarray}
    \alpha\frac{\rmi b}{2\sin\omega_0}
     \ _{p-1}C_\frac{p}{2}
     z_n^{\frac{p}{2}}\overline{z_n}^{\frac{p}{2}-1}
    +\frac{\rmi b}{2\sin\omega_0}\overline{z_n}^{p-1}
    -\rmi\delta z_n
   =0.
   \label{eq:3-20}
\end{eqnarray}
Setting $z_n:=re^{\rmi\theta}$ where $r$ and $\theta\in\mathbb{R}$,
\begin{eqnarray}
 \sin (p\theta)=0,
 \ {\rm and}\ 
 r^{p-2}=\frac{2\delta\sin\omega_0}
               {b\left(
                   \alpha    
                   _{p-1}C_\frac{p}{2}   
                   +\cos (p\theta)
                 \right)}
        >0,
   \label{eq:3-21}
\end{eqnarray}
which reads
\begin{eqnarray}
  &r^{p-2}=\frac{2\delta\sin\omega_0}
                 {b\left(\alpha
                   _{p-1}C_\frac{p}{2}+1
                   \right)}
   \qquad 
   &{\rm for}\ \theta=\frac{\pi}{p}2l,
   \label{eq:3-22} \\
  &r^{p-2}=\frac{2\delta\sin\omega_0}
                 {b\left(\alpha
                    _{p-1}C_\frac{p}{2}-1
                   \right)}
   &{\rm for}\ \theta=\frac{\pi}{p}(2l+1),
   \label{eq:3-23}
\end{eqnarray}
where $l=0,1,\ldots,p-1$. 
When $p$ is even number which satisfies $p\geq4$ and $b\delta>0$, there
are $2p$ fixed points. 
In this case, it can be shown that the half of them are elliptic, while
the others are hyperbolic, and the Poincar\'e-Birkoff bifurcation occurs.  
Here the resonance structure in this case consists of a chain of
$p$ resonant islands.

\section{\label{sec:conclusion}Concluding remarks} 
The renormalization group method with Lie symmetry is developed and
applied to difference equations such as symplectic map.  
When we apply the conventional renormalization group method to a singular
perturbation problem of difference equations, it is difficult to find
how to construct the renormalization group equation, which is a
continuous differential equation. 
In this paper, we succeed in the consistent derivation of a
renormalization group equation to a difference equation by employing the
theory of Lie symmetry. Our results are significant in the
sense that a long-time behavior of a perturbed difference equation is
described by a continuous differential equation, that is the 
renormalization group equation. 
Furthermore, for the case of a simplistic map, it is shown the
renormalization group equation becomes a Hamiltonian system. Then the phase
space structure can be described by the Hamiltonian.
As an application of our theory, we analyze the Poincar\'e-Birkoff
bifurcation in a 2-D symplectic map and the condition of the 
Poincar\'e-Birkoff bifurcation is given to the 2-D symplectic map. 

\section*{Ackowlegdement}
The authors are grateful to Professor D. Levi, Universita' degli Studi di Roma Tre, for fruitful discussions. This research is partially supported by a Grant-in-Aid from Nagoya University 21st Century COE (center of excellence) program "ORIUM".

\end{document}